\documentstyle[prl,aps,epsfig,preprint]{revtex}
\newcommand{\be}{\begin{equation}}
\newcommand{\ee}{\end{equation}}
\newcommand{\bea}{\begin{eqnarray}}
\newcommand{\eea}{\end{eqnarray}}
\newcommand{\beas}{\begin{eqnarray*}}
\newcommand{\eeas}{\end{eqnarray*}}
\newcommand{\bi}{\begin{itemize}}
\newcommand{\ei}{\end{itemize}}

\newcommand{\ba}{\begin{array}}
\newcommand{\ea}{\end{array}}
\begin{document}

\title{Lattice models of disorder with order}
\author{Alberto Petri}
\address{CNR, Istituto "O.M. Corbino",
via del Fosso del Cavaliere 100, 00133 Roma, Italy \\
Unit\`a INFM, Universit\`a  La
Sapienza, p.le Aldo Moro 2, 00185 Roma, Italy} \maketitle
\begin{abstract}
This paper  describes  the use of simple lattice
models for studying the properties of structurally disordered
systems like glasses and granulates.
The models considered  have
crystalline states as ground states, finite connectivity, and are
not subject to constrained evolution rules.  After a short
review of some of these models, the paper discusses how two 
particularly simple kinds of models, the Potts model and the exclusion models, 
evolve after a quench at low temperature to glassy states rather than to 
crystalline states. 

\end{abstract}

\section{Introduction}

Recent years have seen a growing interest in systems
commonly found in disordered structural configurations such as
glasses and dense granulates. Clearly different in many respects,
these two classes of materials share in common the property of
displaying stable states which are very far from what would be expected from
equilibrium considerations  \cite{debenedetti01,jaeger96}. As a
consequence of this property other features emerge, like slow
relaxation and response functions.

Models of disorder are generally based on the presence of quenched,
or \textit{a priori}, disorder which takes the system far from
ordered configurations. Only recently  it has been observed that
lattice models capable of ordering can also well reproduce many
properties of disordered systems. On the other hand glass formers
in nature usually have crystalline states as ground states, but
this does not prevent them from being frequently found in some
glassy state. The situation for a granular system is in general
different, insofar as irregular grains inhibit the evolution of
ordered states. Nevertheless even in the case of identical beads
it is practically impossible to arrange them into an ordered
fashion merely by supplying the energy to them.

The aim of this paper is to briefly review some lattice models
employed in the description of the slow dynamics of glasses and
dense granular matter, which do not possess any {\em a priori} disorder 
nor long
range interactions. We shall not discuss in detail the models, but
just recall some of the main results, addressing the interested
reader to the bibliography.  More attention will be drawn to some
recent observations regarding the condition for the emergence of
glassy phases in some models of this kind.

The paper is structured as follows: in Section II exclusion models
and hard particle models are briefly introduced together with
their employment in irreversible dynamics; Section III is devoted
to their use in the description of dense granular matter and
related results; Section IV describes spin models which have been
recently shown to exhibit glassy states and associated slow
dynamics; finally in Sec. V some new ideas are illustrated on when
and why glassy states are generated in the place of crystalline
states in some lattice models.

\section{Exclusion and hard particle models}

 Flory \cite{flory39} modeled the irreversible deposition of dimers
on a one-dimensional surface by the random sequential adsorption
(RSA) of particles on a lattice. In this model, particles are
placed one at time at randomly chosen positions of a linear chain
of sites (``Hard particles''). Each particle occupies two lattice sites 
and cannot
overlap other particles. Flory was able to show that, despite its
stochastic nature, this process leads, in the limit of an infinite
lattice, to a well defined percentage of covering which he computed
exactly. Soon
after the same model (later also called the car parking model) was
extended to two-dimensional systems and to a variety of particles,
also including desorption and cooperative processes  (see
\cite{evans93,evans97} for review).

Diffusivity, which accounts for particle rearrangement and
relaxation, was also introduced in RSA models giving rise to the
so called  RSAD models (see \cite{bartelt91,nielaba97}).
 In addition, multilayer deposition has been
investigated \cite{nielaba97}. RSAD multilayer
models are the base of the most common lattice models for dense
granular materials, as will be explained below.

Exclusion models represent a slight modification of hard particle
models. In the latter each particle occupies a certain set of
lattice sites that cannot be occupied by other particles. In
exclusion models, each particle occupies one single site, and
excludes a certain neighbour set from being occupied by other
particles. Depending on the exclusion rules,  a
one-to-one correspondence can  sometimes be drawn between the 
two kinds of models.

 Exclusion models have been extensively investigated
 in relation to equilibrium phase transitions
\cite{runnels72}. There identical particles are randomly deposited
or retired from a $d$-dimensional lattice at a rate determined by
the value of a chemical potential; in addition particles can
diffuse along the lattice (in accordance with the exclusion
rules), allowing the system to look for configurations of the
maximum density compatible with the imposed value of the chemical
potential $\mu$. When $\mu$ is taken to infinity the maximum allowed
density in reached with particles arranged in an ordered phase. A
remarkable result of these models is that they are able to
describe both first and second order phase transitions as a
function of the space dimensionality and of the number of excluded
sites \cite{runnels72}. On the contrary, when the
chemical potential in RSAD processes is changed abruptly, 
diffusion is not always
sufficient to drive the system into an ordered phase. 
When this happens the ordering process is generally very
slow, with density  evolving algebraically in time.

\section{Models for granulars}

A process widely investigated in the physics of granular materials
is the compaction, that is the decrease of the volume fraction
(the ratio between the total volume of the constituent grains and
the volume macroscopically occupied by the system) of a spatially
limited system, subject to gravity, under the action of an external
perturbation. The energy supplied by the perturbation (e.g. a
mechanical perturbation like tapping or shaking) must be
comparable with the variation in potential energy required to
rearrange the system structure. The process exhibits rather a rich
behaviour with reversible and irreversible cycles as a function of
the perturbation amplitude \cite{nowak98}, and logarithmic
\cite{knight95} or stretched exponential laws \cite{philippe02} in
the number of tapping.

Lattice models with only the excluded volume constraint have
proven to be good models for dense granular systems. They are all
essentially based on multilayer sequential deposition
\cite{deoliveira98}, and have been so far investigated in $1+1$
dimensions. The lattice is considered to lay in the vertical position.
It is initially prepared by sequentially dropping extended
hard particles from the top. Each particle proceeds downwards until
it hits the bottom of the lattice or a previously deposited particle.
Excluded volume effects prevent the process from
generating ordered configurations and an initial state with many
defects and low volume fraction $\rho$ is usually obtained. Then
an external mechanical perturbation is activated by turning on
upwards diffusion along the lattice meshes for a finite time (the
duration of the tapping) and then allowing only downwards and
horizontal diffusion in order to mimic the relaxation process
under the effect of the gravity field. By periodically turning
upwards diffusivity on and off the effect of multiple tapping
cycles can be investigated, modulating their intensity by changing
the ratio between upwards and downwards-horizontal diffusion.

The so called ``Tetris'' model introduced by Caglioti et al.
\cite{caglioti97} in its original version is based on dimers
placed on a
lattice with meshed tilted of $45^\circ$ and, when tapped, 
shows a logarithmic increase of the
volume fraction, in agreement with \cite{nowak98}.
More generally,  "T" shaped particles have also been
considered,
showing that the model possesses a thermodynamic-like temperature
which is numerically measurable through the fluctuation-dissipation
relations \cite{barrat02} and which coincides with a configurational
pseudo-temperature  proposed for granular matter
\cite{edwards}. The latter has also been measured in
a system of monodisperse hard spheres on lattice \cite{fierro02}.

Dimers placed on lattice with non tilted meshes also exhibit very well
defined scaling laws \cite{fusco01} but with stretched exponential
(the so called Kohlrausch-Williams-Watts law) rather
than logarithmic compaction: \be \label{KWW} \rho (t)=\rho_\infty - \Delta
\rho \exp(-t/\tau_0)^\beta. \ee A similar behaviour 
has been found in recent extensive experiments \cite{philippe02} and 
in a generalization of the model
which includes friction \cite{fusco02}.

\section{Models for glasses}

Lattice gas models with constrained dynamics \cite{kob93}
exhibit many of the properties of glass dynamics, such as slow
relaxation, two step relaxation, vanishing diffusivity, ageing, etc.
Other kinds of models with constrained dynamics
have also been introduced and investigated
yielding similar
results \cite{fredrickson84,newmann99,buhot01,davison01}. 

A class of lattice models exhibiting glassy behaviour with neither
 quenched disorder nor
constrained dynamics are the spin models with  many spin
interaction whose energy can be written as:
\[
E=-\sum_{<i,j,\dots,k>} s_i s_j\dots s_k,
\]
where the sum is extended to a finite number of Ising spins:
$s_i=\pm 1$. A single spin flip Monte Carlo algorithm is generally used
for numerical simulation of the dynamics.
The four spin model \cite{lipowski97} exhibits a discontinuous 
equilibrium transition in three dimensions, but if
quenched quickly enough below the transition temperature, it
enters a very long living metastable state with many of the features of
glassy systems \cite{lipowski00,swift00}. More recently
\cite{franz01} the existence of a dynamic glass transition has been
proved for a three spin model on a bethe hyper-lattice that 
also has a first order transition to a ferromagnetic phase.

Models with geometrical frustration and only two body interaction have been
employed as coarse grained approximation of hard sphere models in the
continuum limit. A dynamical glass transition has been observed in
three-dimensional models with a constraint on the local equilibrium density 
\cite{biroli02} together with a first order transition.
Similar results have also been  found in
\cite{weigt02} and in \cite{pica02} although in the latter model the
crystalline phase does
not seem accessible  by numerical simulation even at very
low cooling rate.

Some kind of glassy states can also be observed in two dimensional
exclusion models. It has been found by numerical simulations
\cite{eisenberg00}
that a monolayer RSAD of particles with the
exclusion of the first three shells of neighbours ($N_3$ model)
is not able to generate a periodic arrangement.
The dynamics leads asymptotically to states in which 
many locally ordered domains of particles are formed, while
at the domain
boundaries many sites remain empty.  Analogous behaviour has been
observed in the RSAD of dimers \cite{grigera97,fusco01a},
whereas in models with exclusion of only two shells of neighbours
($N_2$ model) the density of empty sites vanishes in the
thermodynamic limit \cite{eisenberg98}. In
the next section the properties of these models will be discussed more
extensively.

The many spin interaction models, the $N_3$ model and the other
models referred to above show that glassy states are generated also
in models with no quenched disorder or long range interaction and
with stable crystalline states. On the
other hand, glass formers in general also have  crystalline
phases; this point deserves
consideration and in the next section some ideas on the emergence
of glassy vs crystalline states in some lattice models are discussed.

\section{Emergence of glassy states in lattice models}

In this section some properties are reported that have recently
emerged \cite{deoliveira02a,deoliveira03a,deoliveira03b} in the
coarsening of the Potts model following a quench and in RSAD
dynamics of some exclusion models.

In the Potts model  \cite{wu92}  each lattice site can stay in
one out of $q$ distinct states, or colours. To each site one
associates a variable $\eta_i$ that takes the values
$0,1,2,\dots,q-1$. Let us write the energy of the $q$ state Potts
model as:
\[
U=\sum_{(ij)} (1-\delta_{\eta_i \eta_j}),
\]
where $\eta_i=1,2, \dots q$ and the sum is over the nearest neighbour
sites. There are $q$ degenerate system ground states with zero energy
corresponding to having all sites the same colour.

Starting from an initial disordered configuration  a quench of the
system can be numerically simulated by letting the system evolve
according to the Metropolis dynamics \cite{metropolis53} at a
temperature $T$ below the temperature at which the model is known
to have a  phase transition. For $T=0$, large $q$ and
periodic boundary conditions, the dynamics lead the system into
some frozen state \cite{anderson89} that represents a local
minimum in the energy landscape of the finite system with energy
higher than the ground state energy. This is not the case for low
$q$. In the two-dimensional Ising model, corresponding to $q=2$,
the fraction of blocked configurations tends to zero in the
thermodynamic limit.

A first important consideration concerns such a different behaviour exhibited
for different values of $q$. Figure~1 shows the decrease in time
of the energy per site, $E=U/N$, in a system with $N=10^6$ sites for
some different values of $q$ after a quench at $T=0$. In this case
fixed boundary conditions have been used, that is a prefixed
colour has been assigned at the beginning to the sites at the
system boundary, and cannot change during the simulation.
 This breaks
the symmetry among the different ground states and makes the
system converge to the preselected ground state.
 Since it is expected to
follow an algebraic decay of the type $E(t) \propto t^{-\alpha}$
\cite{anderson89,derrida96}, the curves have been plotted on a
double logarithmic scale. The figure shows that the lines, which
are straight  for $q=2$ and $q=3$, decrease their slope and
develop a bump as $q$ increases.  This suggests at first sight a
deviation from the algebraic behaviour. However if the same curves
are plotted as functions of $1/\sqrt(t)$ a linear behavior is
observed for all times up to the final relaxation for every value
of $q$, showing that the energy decrease in this regime is
correctly described by
\[
E(t) \propto t^{-1/2} + \text{const}
\]
as shown in Fig.~2 for the case $q=7$.
It is the constant term, which  depends on $q$, that accounts
for the bump and bending observed in log-log scale.
After the $1/\sqrt(t)$ regime the system performs a fast relaxation towards
the prefixed ground state.

The second important point is that the time at which the final
relaxation occurs increases with lattice size $N$. The same
happens for the freezing when  periodic boundary conditions are used,
but in this case fluctuations of the final energy from sample to
sample are larger. Extrapolation of the energy curves leads to conclude
that
in the thermodynamic limit:
 \be \label{excess} \lim_{t \rightarrow \infty} E(t) =
E^*(q). \ee It can be useful to point out that here the
thermodynamic limit is taken before the infinite time limit. This
is the reason for which the thermodynamical equilibrium state with
$E=0$ is not obtained.

The residual energy term (\ref{excess}) is due to interfaces
separating homogeneous domains with different values of $q$ that
do not disappear when $t \rightarrow \infty$. Moreover, since for
$q$ large enough this term does not vanish  when $N\rightarrow
\infty$, it can be concluded that the total length of the
interfaces is of order $N$. This property can be used to define the 
glassy
state in this kind of model, and is observed for any $q>4$. For
$q < 4$  the residual energy vanishes in the thermodynamic limit,
showing absence of interfaces (crystalline state for $q=2$) or at
most a total length of order $\sqrt(N)$ (polycrystalline state for
$q=3$). The case $q=4$ is critical insofar as relaxation is not
$1/\sqrt t$; by including a logarithmic correction $E(t) \propto
\sqrt (\ln t)/t + E^*$ it is compatible with $E^*=0$.

Simulations demonstrate that the residual energy $E^*(q)$ is a
monotonic function of $q$. This is shown for some values of $q$ in
Fig.~3 together with the fitting curve $E^*(q)=(q-q_c)^{0.5}$ for
$q \ge q_c=4$. This suggests a relationship between a non zero
residual energy and the discontinuous transition in the
equilibrium phase diagram of the model displayed when $q>4$. Such
a conjecture is made stronger by the observation of an analogous
behaviour in the case of exclusion models. They are predicted to
belong to the same universality class as the Potts model
\cite{domany77}, with a correspondence between the number $p$ of
excluded sites per particle and  $q$. In the $N_1$ model $p=2$,
whereas $p=4$ and $p=5$ in the $N_2$ and $N_3$ models
respectively. In these systems the density of deposited particles
plays a role similar to that of energy in the Potts model; in fact
one can associate an energy to the density by means of the
chemical potential $\mu$.  In particular, the density is
maximum, and the energy is minimum, in the state with the 
closest allowed packing of particles, corresponding to a periodic
arrangement (ground state). Any deviation from this value signals
the presence of residual empty sites (defects). By assuming 
$E=0$ for the state with the closest packing, 
an energy  $E=\mu \rho_d$ corresponds to a density of defects $\rho_d$.
 The equilibrium
diagrams of these systems exhibit continuous transition for $p=2$
and $p=4$ and discontinuous transition for $p=5$
\cite{runnels72,orban82,domany77}. Thus zero density of defects is
expected for $p=2$ (crystalline states),
 a vanishing one  for $p=3,4$  (polycrystalline states) and glassy states
with a number of defects $O(N)$ for $p=5$. This has indeed been
observed \cite{eisenberg00,eisenberg98,wang93a} and has been
confirmed by further numerical simulations to be largerly
independent on the chosen parameters
\cite{deoliveira03a,deoliveira03b}. On a triangular lattice $N_1$
shows an equilibrium first order transition, whereas the
transition is discontinuous for the $N_1$ and the $N_2$ models.
 Numerical simulations confirm the prediction of crystalline states for
$N_1$ but are not yet conclusive for $N_2$ and $N_3$
\cite{deoliveira03a}. It can be interesting to mention that
a kind of polycrystalline state, in the sense defined above, has also been
observed in the quench of the three-dimensional Ising model
\cite{spirin02}.

\section{Summary}

The work done in recent years has shown that lattice models with
crystalline  states and finite connectivity are capable of
reproducing many properties of structurally disordered systems
like granulates and glasses. This is of relevance for the
understanding of the underlying mechanisms, also in view of
similar behaviours exhibited by other kinds of models whose 
features  can however be traced back to different origin \cite{franz01a}. 

The analysis of the conditions under which a lattice model with
crystalline states may exhibit glassy states has been recently
carried out on simple models such as exclusion and Potts models
showing that it is the competition among a number of equivalent
ground states that drives the systems away from crystallisation.
When this number is small crystallisation or polycrystallisation
take place instead.  This result is obtained by inverting the
usual order of performing the infinite time and the thermodynamic 
limits, and suggests a relationship between the observed behaviour and the
equilibrium properties of the systems, with  glassy behaviour
associated to the presence of a first order transition which is 
generallly also a feature of the other models
described in this paper.

Many thanks are due to T. Tom\'{e} and M. J. de Oliveira, with whom
most of the work described in Sec. V has been developed, and to F.
Dalton for revising  the manuscript.

\begin{figure}
\centering\epsfig{file=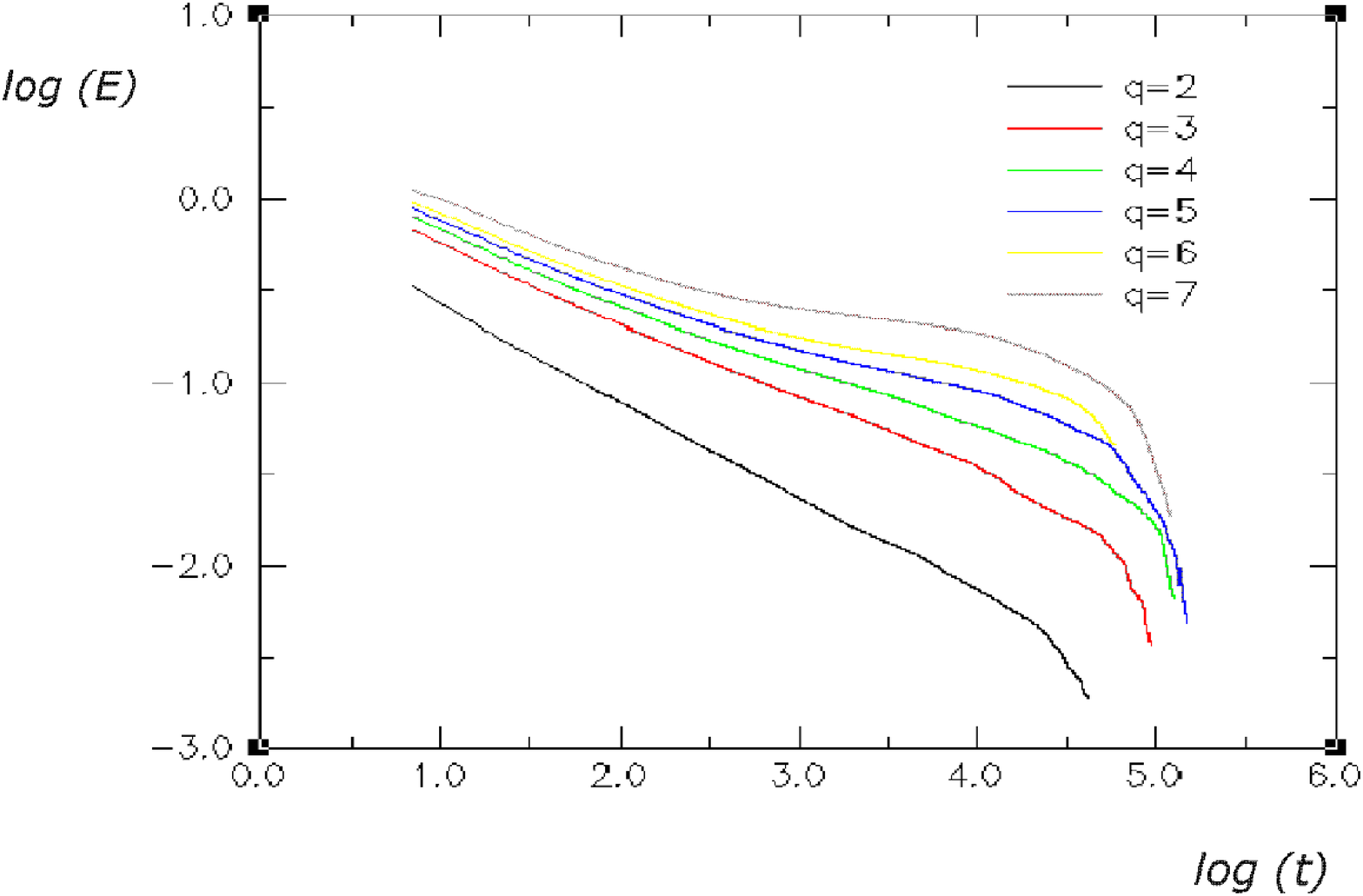, width=12cm, height=9cm}
\caption{Energy decay in a two dimensional Potts model after a
quench at $T=0$. For increasing $q$ larger deviations from $t
\propto 1/\sqrt(t)$ are displayed.}
\end{figure}

\begin{figure}
\centering\epsfig{file=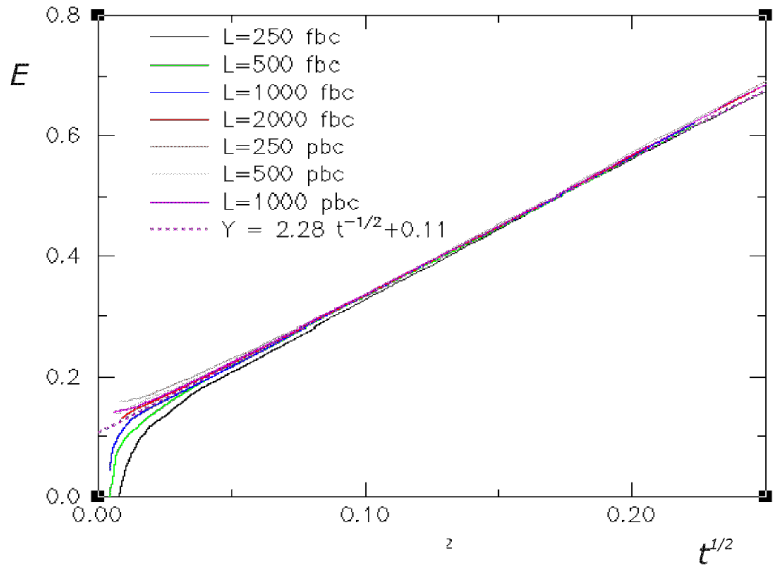, width=12cm, height=9cm}
\caption{E(t) vs $1./\sqrt(t)$ for the Potts model with  $q=7$ and
different size and boundary conditions: {\em fbc}= fixed boundary
conditions; {\em pbc} = periodic boundary conditions. A common regime is
shown whose duration increases with the system size.}
\end{figure}

\begin{figure}
\centering\epsfig{file=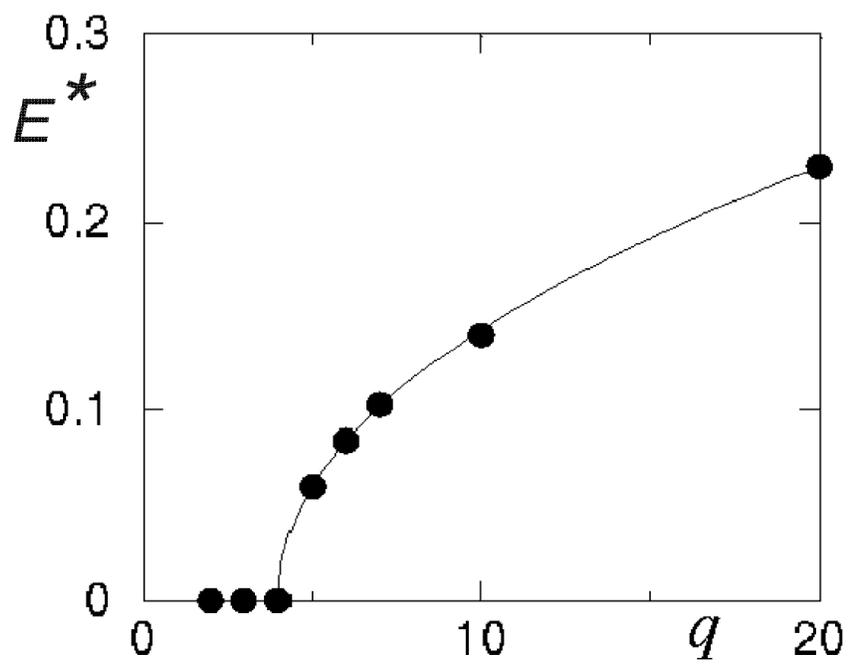, width=12cm, height=10cm}
\caption{Residual energies in the Potts model after a
quench at $T=0$ as a function of the number of states $q$.}
\end{figure}

\end{document}